\documentclass{PoS}
\usepackage{cite}
\usepackage{amsmath}
\usepackage[table]{xcolor}

\newcommand{\yellow}{\cellcolor[HTML]{F9E79F}}
\newcommand{\green}{\cellcolor[HTML]{ABEBC6}}

\newcommand{\blue}{\cellcolor[HTML]{AED6F1}}
\newcommand{\dblue}{\cellcolor[HTML]{E8DAEF}}
\newcommand{\purple}{\cellcolor[HTML]{F5B7B1 }}
\newcommand{\red}{\cellcolor[HTML]{F2D7D5}}
\newcommand{\orange}{\cellcolor[HTML]{E59866}}
\newcommand{\M}{\mathcal{M}}
  \newcommand{\Mtree}{\M^{\mathrm{(tree)}}}

\newcommand{\Wf}{\Omega}
  \newcommand{\Wfl}[1]{\Wf^{({#1})}}

\def\nn{\nonumber}
\newcommand{\El}{\mathcal{L}}

  \newcommand{\Wtdl}[1]{\Wfl{#1}_{\mathrm{2d}}}
\newcommand{\zb}{\overline{z}}
\newcommand{\wb}{\overline{w}}

\newcommand{\Dd}{\mathrm{D}}

  \newcommand{\Dk}{[\Dd k]}
  
\newcommand{\dd}{\mathrm{d}}

  \newcommand{\ddeps}{\dd^{2-2\epsilon}}

\newcommand{\Ha}{\hat{H}}

  \newcommand{\Htd}{\Ha_{\mathrm{2d}}}
	\newcommand{\Hitd}{\Ha_{\mathrm{2d,i}}}
	\newcommand{\Hmtd}{\Ha_{\mathrm{2d,m}}}

\newcommand{\T}{\mathbf{T}}
  \newcommand{\Ts}{\T_s^2} 
  \newcommand{\Tt}{\T_t^2} 
  \newcommand{\Tu}{\T_u^2} 
  \newcommand{\Tsu}{\T_{s-u}^2}
\newcommand{\Ca}{C_A}

\newcommand{\Cone}{(2\Ca-\Tt)}
  \newcommand{\Ctwo}{(\Ca-\Tt)}

\def\T{{\bf T}}
\def\Tt{{\bf T}_t^2} 
\def\Ts{{\bf T}_s^2} 
\def\Tu{{\bf T}_u^2} 
\def\Tsu{{\bf T}_{s{-}u}^2}

\title{The High-Energy Limit of 2-to-2 Partonic Scattering Amplitudes}

\ShortTitle{QCD amplitudes in the Regge limit}

\author{\speaker{Einan Gardi}\\
        Higgs Centre for Theoretical Physics, 
School of Physics and Astronomy, \\
The University of Edinburgh, Edinburgh EH9 3FD, Scotland, UK\\
        E-mail: \email{Einan.Gardi@ed.ac.uk}}

\author{Simon Caron-Huot\\
Department of Physics, McGill University, 3600 rue University, Montr\'eal, QC Canada H3A 2T8\\
        E-mail: \email{schuot@physics.mcgill.ca}}

\author{Joscha Reichel\\
       Higgs Centre for Theoretical Physics, 
School of Physics and Astronomy, \\
The University of Edinburgh, Edinburgh EH9 3FD, Scotland, UK\\
        E-mail: \email{joscha.reichel@ed.ac.uk}}

\author{Leonardo Vernazza\\
       Nikhef, Science Park 105, NL--1098 XG Amsterdam, The Netherlands\\
        E-mail: \email{l.vernazza@nikhef.nl}}

\abstract{Recently, there has been significant progress in computing scattering amplitudes in the high-energy limit using rapidity evolution equations.  We describe the state-of-the-art and demonstrate the interplay between exponentiation of high-energy logarithms and that of infrared singularities. 
The focus in this talk is the imaginary part of 2 to 2 partonic amplitudes, which can be determined by solving the BFKL equation. We demonstrate that the wavefunction is infrared finite, and that its evolution closes in the soft approximation. Within this approximation we derive a closed-form solution for the amplitude in dimensional regularization, which fixes the soft anomalous dimension to all orders at NLL accuracy. 
We then turn to finite contributions of the amplitude and show that the remaining hard contributions can be determined algorithmically, by iteratively solving the BFKL equation in exactly two dimensions within the class of single-valued harmonic polylogarithms.
To conclude we present numerical results and analyse large-order behaviour of the amplitude.}

\FullConference{14th International Symposium on Radiative Corrections (RADCOR2019)\\ 
9-13 September 2019\\
		Palais des Papes, Avignon, France}

\begin{document}

\section{Introduction}

The high-energy (Regge) limit\footnote{For a recent introduction to the general subject of scattering amplitudes in the high-energy limit see \cite{White:2019ggo}.} of QCD scattering has been an active area of research for many years e.g.~\cite{Kuraev:1977fs,Balitsky:1978ic,Lipatov:1985uk,Mueller:1993rr,Mueller:1994jq,Brower:2006ea,Moult:2017xpp}.
While the general problem of high-energy scattering is non-perturbative, in the regime where the exchanged momentum $-t$ is high enough,  i.e. $s\gg-t\gg\Lambda_{\rm QCD}^2$, perturbation theory offers systematic tools to analyse this limit. Central to this is the Balitsky-Fadin-Kuraev-Lipatov (BFKL) 
evolution equation~\cite{Kuraev:1977fs,Balitsky:1978ic}, which provides a systematic theoretical framework to resum high-energy (or rapidity) logarithms, $\ln (s/(-t))$, to all orders in perturbation theory. 
It was used extensively to study a range of physical phenomena including the small-$x$ limit of parton density functions and jet production at large rapidity. Furthermore, the non-linear generalisation of BFKL, the Balitsky-JIMWLK equation~\cite{Balitsky:1995ub,Balitsky:1998kc,Kovchegov:1999yj,JalilianMarian:1996xn,JalilianMarian:1997gr,Iancu:2001ad}, 
is today a main tool in the study of heavy-ion collisions.

While direct application of rapidity evolution equations to phenomenology requires the scattering particles to be colour-singlet objects, here we are concerned with the theoretical problem of understanding  \emph{partonic} scattering amplitudes in the high-energy limit, similarly to refs.~\cite{Sotiropoulos:1993rd,Korchemsky:1993hr,Korchemskaya:1996je,Korchemskaya:1994qp,DelDuca:2001gu,DelDuca:2013ara,DelDuca:2014cya,Bret:2011xm,DelDuca:2011ae,Caron-Huot:2013fea,Caron-Huot:2017fxr,Caron-Huot:2017zfo}. 
This is part of a more general programme of understanding the structure of gauge-theory amplitudes and the underlying principles governing this structure. Taking the high-energy limit results in a drastic simplification of the dynamics as compared to general kinematics, a simplification that is captured by the above-mentioned rapidity evolution equations.
By making use of this, amplitudes may be computed to all orders, to a given logarithmic accuracy. 

In a recent series of papers~\cite{Caron-Huot:2013fea,Caron-Huot:2017fxr,Caron-Huot:2017zfo} we have studied $2\to 2$ partonic amplitudes, $qq\to qq$, $gg\to gg$, $qg\to qg$, in QCD and related gauge theories by making use of rapidity evolution equations.
In ref.~\cite{Caron-Huot:2017fxr} we focused on the real part of the amplitude through three loops and determined the effects associated with the exchange of three Reggeised gluons by making use of the Balitsky-JIMWLK equation.
The present talk, based primarily on ref.~\cite{Caron-Huot:2017zfo} and on an upcoming publication \cite{Caron-Huot:2019_TBP}, focusses on the imaginary part of these $2\to 2$ amplitudes, where high-energy logarithms are generated by the exchange of a pair of Reggeised gluons. In these papers we were able to take a major step forward and compute the first tower of high-energy logarithms by iteratively solving the BFKL equation.

The outline of the talk is as follows. In section~\ref{sec:Regge} we provide a brief introduction to the topic of $2\to2$ partonic scattering amplitudes in the high-energy (Regge) limit and introduce the concept of signature and its relation to colour flow. In section \ref{sec:IR} we present the complementarity between the study of this limit and the study of infrared singularities. Next, in section \ref{sec:BFKL_NLL} we focus on the imaginary part of $2\to2$ partonic amplitudes and briefly present the BFKL equation for a general colour exchange in dimensional regularization. In section~\ref{sec:BFKL_soft} we review the approach of \cite{Caron-Huot:2017zfo} where the BFKL equation was considered in the soft limit and used to compute the infrared-singular part of the first tower of high-energy logarithms contributing to the imaginary part of the amplitude.
In section~\ref{sec:BFKL_2d} we proceed to review new results, to be published in~\cite{Caron-Huot:2019_TBP}, regarding the finite parts of the same tower of corrections by making use of BFKL evolution in strictly two transverse dimensions. In section~\ref{sec:results} we discuss our results, emphasising in particular the large order behaviour of the series for both the anomalous dimension and the finite corrections to the amplitude. In section~\ref{sec:conclusions} we briefly summarise our conclusions.

\section{The high-energy limit in $2\to 2$ partonic scattering\label{sec:Regge}}

Scattering amplitudes of quarks and gluons are dominated 
at high energies by the $t$-channel exchange of effective degrees of freedom called \emph{Reggeized gluons}.  $2\to 2$ amplitudes are conveniently decomposed into \emph{odd} and \emph{even} signature characterising their symmetry properties under $s\leftrightarrow u$ interchange:
\begin{equation}\label{Odd-Even-Amp-Def}
 {\cal M}^{(\pm)}(s,t) = \tfrac12\Big( {\cal M}(s,t) \pm {\cal M}(-s-t,t) \Big)\,,
\end{equation}
where odd (even) amplitudes ${\cal M}^{(-)}$ (${\cal M}^{(+)}$) are governed by the exchange of an odd (even) number of Reggeized gluons. Furthermore, as shown 
in ref.~\cite{Caron-Huot:2017fxr}, these have respectively 
\emph{real} and \emph{imaginary} coefficients, when expressed 
in terms of the natural signature-even combination of 
logarithms,
\begin{equation}
\label{L-def}
\frac12\left(\log\frac{-s-i0}{-t}+\log\frac{-u-i0}{-t}\right)
\simeq \log\left|\frac{s}{t}\right| -i\frac{\pi}{2} \equiv L\,.
\end{equation}

Owing to Bose symmetry, 
the symmetry of the colour representation of the $t$-channel exchange mirrors the signature of the corresponding amplitude coefficients. Considering e.g. gluon-gluon scattering, the full set of possible representations
$8 \otimes 8 = 1 \oplus 8_{s} \oplus 8_{a} 
\oplus 10 \oplus \overline{10} \oplus 27 \oplus 0$
is naturally separated into signature odd and even:
\begin{equation} \label{odd_even_Ms}
\mbox{odd: } {\cal M}^{[8_a]}, {\cal M}^{[10+\overline{10}]},
\qquad
\mbox{even: } 
{\cal M}^{[1]},
{\cal M}^{[8s]}, 
{\cal M}^{[27]},
{\cal M}^{[0]} \qquad \mbox{($gg$ scattering)\,.}
\end{equation}

The real part of the amplitude, ${\cal M}^{(-)}$, is governed, at leading logarithmic (LL) accuracy, by the exchange of a single Reggeized gluon in the $t$ channel. To this accuracy, high-energy logarithms admit a simple exponentiation pattern, namely 
\begin{equation}
\label{Mreal}
{\cal M}^{(-)}_{\rm LL} = (s/(-t))^{\alpha_g(t)} \times \Mtree
\end{equation}
where the exponent is the \emph{gluon Regge trajectory} (corresponding to a Regge pole in the complex angular momentum plane),
$\alpha_g(t)=\frac{\alpha_s}{\pi} C_A \alpha_g^{(1)}(t)+{\cal O}(\alpha_s^2)$, whose leading order coefficient $\alpha_g^{(1)}(t)$ is infrared singular, $\alpha_g^{(1)}(t)\sim \frac{1}{2\epsilon}$ in dimensional regularization with $d=4-2\epsilon$.  $\Mtree$ represent the tree amplitude, 
given by
\begin{equation}
  \Mtree_{ij\to ij} = 4\pi \alpha_s \frac{2s}{t} (T_i^b)_{a_1 a_4} (T_j^b)_{a_2 a_3} 
  \delta_{\lambda_1\lambda_4}\delta_{\lambda_2\lambda_3},
\end{equation}
where $\lambda_i$ for $i=1$ through $4$ are helicity indices and $T_i^b$ and $T_j^b$ are the colour matrices in the representations of the scattered partons.

In this talk our main interest is the imaginary part of the amplitude, ${\cal M}^{(+)}$. Here the leading tower of logarithms, on which we focus in sections~\ref{sec:BFKL_NLL} through \ref{sec:results} below, is generated by the exchange of \emph{two} Reggeized gluons, starting with a non-logarithmic term at one loop:
\begin{equation}
\label{MevenOneloop}
{\cal M}^{(+)}_{\rm NLL}\simeq 
i\pi 
\left[\frac{1}{2\epsilon} \frac{\alpha_s}{\pi}+{\cal O}\left(\alpha_s^{2} L\right)\right]{\mathbf T}^2_{s-u}  {\cal M}^{\rm tree}\,.
\end{equation}
Here we suppressed subleading terms in $\epsilon$ as well as multiloop corrections, which take the form $\alpha_s^l L^{l-1}$ at $l$ loops; because the power of the energy logarithm $L$ is one less than that of the coupling, these are formally next-to-leading logarithms (NLL).   

The colour operator $\Tsu$ in eq.~\eqref{MevenOneloop} acts 
on $\Mtree_{ij\to ij}$ and it is defined in terms of the usual 
basis of quadratic Casimirs corresponding to colour flow through 
the three channels~\cite{Dokshitzer:2005ig,DelDuca:2011ae}:
\begin{equation}
  \label{TtTsTu}
  \Tsu \equiv \frac{\Ts-\Tu}{2} \qquad \rm{with} \qquad \left\{ \begin{array}{c}
    \T_s = \T_1+\T_2=-\T_3-\T_4, \\ 
    \T_u = \T_1+\T_3=-\T_2-\T_4, \\
    \T_t = \T_1+\T_4=-\T_2-\T_3,
  \end{array} \right.
\end{equation}
where $\T_i$ is the colour-charge operator 
\cite{Catani:1998bh} associated with parton $i$.  
In eq.~(\ref{MevenOneloop}) one may observe that the colour structure is even under $s\leftrightarrow u$ interchange: ${\cal M}^{\rm tree}$ is odd, and so is the operator ${\mathbf T}^2_{s-u}$ acting on it.

\section{Complementarity between the high-energy limit and long-distance singularities\label{sec:IR}}

A central research direction in understanding the structure of perturbative gauge-theory amplitudes is the study of their long-distance (infrared) singularities~\cite{Korchemsky:1993hr,Korchemskaya:1996je,Korchemskaya:1994qp,Catani:1996vz,Catani:1998bh,Sterman:2002qn,Dixon:2008gr,Kidonakis:1998nf,Bonciani:2003nt,Dokshitzer:2005ig,Aybat:2006mz,Gardi:2009qi,Becher:2009cu,Becher:2009qa,Gardi:2009zv,
  Dixon:2009gx,Dixon:2009ur,Bret:2011xm,DelDuca:2011ae,Caron-Huot:2013fea,Ahrens:2012qz,Naculich:2013xa,Erdogan:2014gha,Gehrmann:2010ue,Falcioni:2019nxk,Becher:2019avh}. This concerns amplitudes in general kinematics, not specifically the high-energy limit, but as we shall see there is an interesting interplay between the study of infrared singularities and of the high-energy limit.
This relation is deep and goes back a long way~\cite{Sotiropoulos:1993rd,Korchemsky:1993hr,Korchemskaya:1994qp,Korchemskaya:1996je,DelDuca:2001gu,DelDuca:2013ara,DelDuca:2014cya,Bret:2011xm,DelDuca:2011ae,Caron-Huot:2013fea,Caron-Huot:2017fxr,Caron-Huot:2017zfo}; we will only illustrate here certain aspects it entails.
 
Much progress have been achieved in recent years in studying infrared singularities in massless multi-leg amplitudes at the multi-loop level. It was established a decade ago~\cite{Catani:1998bh,Aybat:2006mz,Becher:2009cu,Gardi:2009qi,Becher:2009qa,Gardi:2009zv} that the soft anomalous dimension governing infrared singularities in multi-leg scattering amplitudes has a remarkably simple structure through two loops: it is given by a sum over colour dipoles which involves only pairwise interactions amongst the hard partons. It was shown then that higher-order corrections that involve multi-leg interactions may start at three loops, and depend on the kinematics via conformally-invariant cross ratios $\rho_{ijkl} = \frac{(-s_{ij})(-s_{kl})}{(-s_{ik})(-s_{jl})}$, so the anomalous dimension takes the form:
\begin{eqnarray} 
\label{gammaSoft1}
&&{\Gamma}_{n}\left(\{p_i\},\lambda, \alpha_s(\lambda^2) \right) \,=\,
{\Gamma}_{n}^{\rm dip.}\left(\{p_i\},\lambda, \alpha_s(\lambda^2) \right)
\,+\,{\Delta}_{n}\left(\{\rho_{ijkl}\}\right)
\\ && \nonumber\\
&\qquad\text{with}&\qquad \nonumber
{\Gamma}_{n}^{\rm dip.}\left(\{p_i\},\lambda, \alpha_s(\lambda^2) \right)=
-\frac{\widehat{\gamma}_{K}  (\alpha_s)}{2} \, \, \sum_{i<j} 
\log \left(\frac{-s_{ij}}{\lambda^2}\right) \, \T_i \cdot \T_j \,+\, \sum_i \gamma_i (\alpha_s) \,,
\end{eqnarray}
where the sum extends over all hard coloured partons, $\widehat{\gamma}_K$ is the cusp anomalous dimension, stripped of the colour Casimir $C_i$, and $\gamma_i$ is the collinear anomalous dimension.
The three-loop corrections $\Delta^{(3)}$ for any number of legs $n$ was computed diagrammatically in ref.~\cite{Almelid:2015jia}. Despite the complexity of the calculation the result is remarkably simple; we refer the reader to the original publication for details. 
Subsequently, this result was reproduced (up to a single rational normalization factor) by a so-called bootstrap procedure~\cite{Almelid:2017qju}, starting with a general ansatz in terms of the relevant type of functions (single-valued harmonic polylogarithms~\cite{Brown:2004ugm,Brown:2013gia,Schnetz:2013hqa,Pennington:2012zj,Dixon:2012yy,DelDuca:2013lma}) and imposing on it all of the known constraints based on the theory of soft gluon exponentiation~\cite{Gardi:2013ita}, symmetries, and information from special kinematic limits. This work was reviewed in the 2017 edition of the present conference~\cite{Gardi:2018arz}.
\begin{figure}[htb]
\begin{center}
     \includegraphics[width=.75\textwidth]{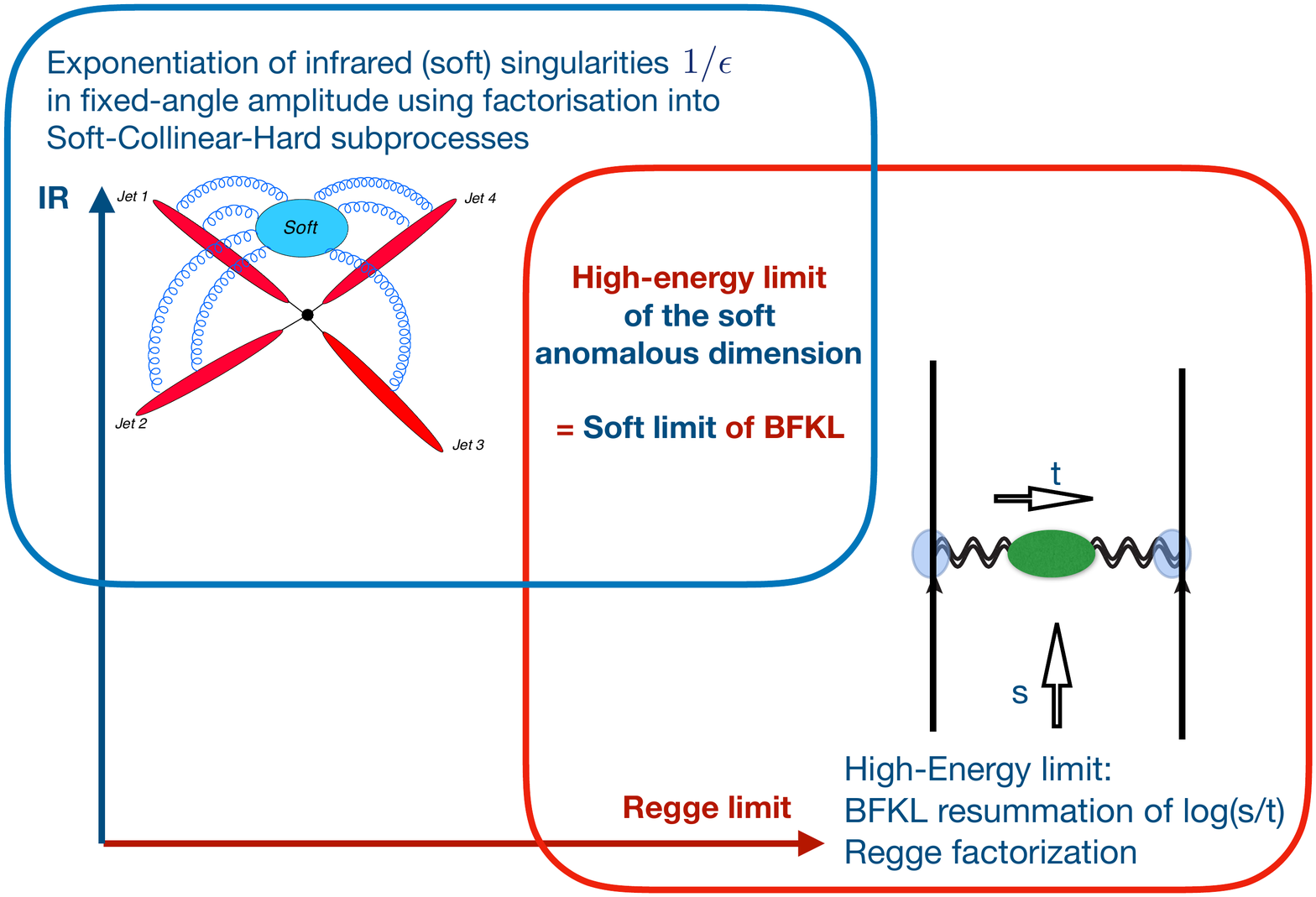}
     \caption{A schematic representation of the two pictures of factorization and exponentiation: soft-gluon exponentiation in fixed angle scattering and Regge exponentiation in the high-energy limit.}
     \label{fig1}
\end{center}
\end{figure}

Chief amongst the constraints imposed on the ansatz for the soft anomalous dimension in~\cite{Almelid:2017qju} was its high-energy limit. This is a prime example of the complementarity of the two limits, which is illustrated in fig.~\ref{fig1}, and will be briefly explained below.
The starting point for analysing infrared singularities is a factorization of soft and collinear singularities in general kinematics, i.e. assuming fixed-angle scattering where all hard scales are simultaneously taken large compare to the QCD scale. In contrast, resummation of the dominant logarithmic corrections in the high-energy limit relies on factorization in rapidity between the target and projectile. Despite starting with different kinematic configurations, one may connect the two pictures by taking the two limits sequentially: on the one hand one may start by considering soft singularities at fixed angles (the vertical axis in fig.~\ref{fig1}), show that these exponentiate in term of the soft anomalous dimension $\Gamma$ of eq.~(\ref{gammaSoft1}), and then gradually take $s\gg -t$, and isolate leading and subleading towers of logarithms within the soft anomalous dimension:
$\Gamma(\alpha_s) =\Gamma_{\rm LL}+\Gamma_{\rm NLL}+
\Gamma_{\rm NNLL}+\ldots\,.$
On the other hand one may start by considering the forward limit $s\gg -t$ (the horizontal axis in fig.~\ref{fig1}), perform Regge factorization and resum leading and subleading towers of high-energy logarithms $L$ (defined in (\ref{L-def})) using the BFKL equation or its generalizations. Within each of the towers one would encounter infrared-singular as well as finite corrections. The former must originate from soft virtual gluons in loops, and hence may be evaluated by considering the soft limit of BFKL~\cite{Caron-Huot:2017zfo}. 
Of course, ultimately, as illustrated in fig.~\ref{fig1}, the two pictures overlap: the soft limit of BFKL must be consistent with the high-energy limit of the soft anomalous dimension\cite{Bret:2011xm,DelDuca:2011ae,Caron-Huot:2013fea,Caron-Huot:2017fxr,Caron-Huot:2017zfo}. 

The simplest example is provided by gluon Reggeization, which amounts to the exponentiation of leading logarithms according to eq.~(\ref{Mreal}). The exponent, governed by the gluon Regge trajectory $\alpha_g(t)$, is infrared singular. Infrared singularities of course exponentiate independently of the high-energy limit. Thus, one may deduce~\cite{Korchemsky:1993hr,Korchemskaya:1994qp,Korchemskaya:1996je} the structure of this exponent also by considering the general pattern of exponentiation of soft singularities governed by the soft anomalous dimension~(\ref{gammaSoft1}). Upon comparing the two pictures of exponentiations  one indeed finds consistency, where the (infrared singular part of the) gluon Regge trajectory admits
\begin{equation}
\label{alpha_g}
\alpha_g(t)=\frac14 \,\mathbf{T}_t^2\int_0^{-t}\frac{d\lambda^2}{\lambda^2}\widehat{\gamma}_K(\alpha_s(\lambda^2,\epsilon))\,.
\end{equation}

The relation between the two exponentiation pictures has been extended to higher orders and logarithmic accuracy. This extension is non-trivial, and reveals a range of interesting phenomena. 
For the real part of the amplitude, the simple exponentiation pattern generated by a single Reggeized gluon  according to eq.~(\ref{Mreal}) is preserved at the next-to-leading logarithmic (NLL) accuracy, except that it requires ${\cal O}(\alpha_s^2)$ corrections to the Regge trajectory and also  the introduction of ($s$-independent) impact factors, accounting for the interaction of the Reggeized gluon with the target and projectile. Beyond NLL accuracy,  this simple picture breaks down due to multiple 
Reggeized gluon exchange, which form Regge cuts. 
This was demonstrated in ref.~\cite{Caron-Huot:2017fxr}, where these effects were computed through three-loops, by 
constructing an iterative solution of the Balitsky-JIMWLK equation, describing the evolution of three Reggeized gluons and their mixing with a single Reggeized gluon.

The state-of-the-art knowledge of the soft anomalous dimension in the high-energy limit is summarised in tables~\ref{tabReal} and \ref{tabIm}. The tables present the coefficients $\Gamma^{(\ell,k)}$ of $(\alpha_s/\pi)^{\ell} L^k$ of the real and imaginary parts of the soft anomalous dimension, respectively.  Considering table~\ref{tabReal} one immediately observes the remarkable fact the dipole formula ${\Gamma}^{\rm dip.}$ in (\ref{gammaSoft1}) alone, contributing to the first two columns in table~\ref{tabReal}, actually dictates much its structure (see eq.~(4.10) in ref.~\cite{Caron-Huot:2017fxr}). 
This can be understood as follows. The simple exponentiation pattern due to gluon Reggeization, i.e.~(\ref{Mreal}) with (\ref{alpha_g}), implies that the anomalous dimension is one-loop exact at LL level and two-loop exact at NLL level. This explains the sequence of zeros on the two rightmost diagonals in table~\ref{tabReal}.
The vanishing of the NNLL correction for the real part of the anomalous dimension, ${\Delta}^{(+,3,1)}=0$, was deduced both  from the diagrammatic calculation of the three-loop soft anomalous dimension and from the JIMWLK-based analysis of the Regge limit in ref.~\cite{Caron-Huot:2017fxr}. The vanishing of this coefficient, in turn, was essential for recovering the three-loop anomalous dimension via bootstrap. This illustrates very clearly the complementary nature of these two avenues of studying the amplitude.
Non-vanishing beyond-dipole corrections were deduced in ref.~\cite{Caron-Huot:2017fxr} from the explicit calculation of $\Delta^{(3)}$ in~\cite{Almelid:2015jia}, and they are given by 
\begin{align}
\label{DeltaCoef}
\begin{split} 
{\Delta}^{(-,3,1)} =\,& i \pi \, [\Tt,[\Tt, \Tsu]] \frac{1}{4}  \zeta_3 \\
{\Delta}^{(-,3,0)} =\,& i \pi \, [\Tt,[\Tt, \Tsu]] \frac{11}{4} \zeta_4\\
{\Delta}^{(+,3,0)} =\,& \frac{1}{4}  [\Tsu,[\Tt, \Tsu]] \bigg[\zeta_5  - 4 \zeta_2 \zeta_3 \bigg] 
-\frac{\zeta_5+2\zeta_2\zeta_3}{8} \bigg\{ -\frac{5}{8} C_A^2 \Tt 
\\&+
\bigg[ \{\T_t^a, \T_t^d \}  \Big(\{\T_{s-u}^b, \T_{s-u}^c \} 
+ \{\T_{s+u}^b, \T_{s+u}^c \} \Big) 
+\,  
 \{\T_{s-u}^a, \T_{s-u}^d \}
 \{\T_{s+u}^b, \T_{s+u}^c \} \bigg] f^{abe}f^{cde} \bigg\},
\end{split}
\end{align}
where ${\Delta}^{(\pm,\ell,k)}$ represents a correction with $+(-)$ even (odd) signature, at order $(\alpha_s/\pi)^\ell\, L^k$, and where   we defined the colour operator
$\T^a_{s\pm u} \equiv \tfrac{1}{\sqrt{2}} \left(\T^a_s\pm \T^a_u\right)$.

\begin{table}[htb]
\begin{center}
     \begin{tabular}{|c||l|c|c|c|c|c|c|}
\hline
 &$L^0$&$L^1$&$L^2$&$L^3$&$L^4$&$L^5$&$L^6$\\\hline\hline
 $\alpha_s^1$&\blue $\frac{1}{4}\widehat\gamma_K^{(1)}\ln\frac{-t}{\lambda^2}\sum_{i=1}^4 C_i+\sum_{i=1}^4\gamma_i^{(1)}$&\dblue$\frac{1}{2}\widehat\gamma_K^{(1)}\mathbf{T}_t^2$&&&&&\\\hline
 $\alpha_s^2$&\green  $\frac{1}{4}\widehat\gamma_K^{(2)}\ln\frac{-t}{\lambda^2}\sum_{i=1}^4C_i+\sum_{i=1}^4\gamma_i^{(2)}$&\blue $\frac{1}{2}\widehat\gamma_K^{(2)}\mathbf{T}_t^2$&\dblue $0$&&&&\\\hline
 $\alpha_s^3$&\yellow$\frac{1}{4}\widehat\gamma_K^{(3)}\ln\frac{-t}{\lambda^2}\sum_{i=1}^4 C_i+\sum_{i=1}^4\gamma_i^{(3)}+{\Delta}^{(+,3,0)}$&\green $\frac{1}{2}\widehat\gamma_K^{(3)}\mathbf{T}_t^2$ &\blue$0$&\dblue$0$&&&\\\hline
 $\alpha_s^4$&\red&\yellow&\green&\blue$0$&\dblue $0$&&\\\hline
 $\alpha_s^5$&\purple&\red&\yellow&\green&\blue$0$&\dblue $0$&\\\hline
 $\alpha_s^6$&\orange&\purple&\red&\yellow&\green&\blue$0$&\dblue$0$\\\hline
     \end{tabular}
\end{center}
     \caption{The signature even part (real part) of the soft anomalous dimension in the high-energy limit. The coloured cells represent contributions with a fixed logarithmic accuracy: each tower of logarithms has a unique colour.  The 3-loop coefficient ${\Delta}^{(+,3,0)}$ determined from the soft anomalous dimension beyond-dipole correction is given in (\ref{DeltaCoef}).
The unfilled cells at the bottom, starting at ${\cal O}(\alpha_s^4)$ and NNLL, are yet unknown.}
     \label{tabReal}
     \end{table}

\begin{table}[htb]
\begin{center}
     \begin{tabular}{|c||l|c|c|c|c|c|c|}
\hline
 &$L^0$&$L^1$&$L^2$&$L^3$&$L^4$&$L^5$&$L^6$\\\hline\hline
 $\alpha_s^1$&\blue$\frac{1}{2}\widehat\gamma_K^{(1)}i\pi \mathbf{T}_{s-u}^2$ &\dblue 0&&&&&\\\hline
 $\alpha_s^2$&\green$\frac{1}{2}\widehat\gamma_K^{(2)}i\pi \mathbf{T}_{s-u}^2$&\blue$0$&\dblue$0$&&&&\\\hline
 $\alpha_s^3$&\yellow$\frac{1}{2}\widehat\gamma_K^{(3)}i\pi \mathbf{T}_{s-u}^2+{\Delta}^{(-,3,0)}$&\green
${\Delta}^{(-,3,1)} $
&\blue $0$&\dblue$0$&&&\\\hline
 $\alpha_s^4$&\red&\yellow&\green&\blue${\Gamma}^{(-,4)}_{\rm NLL}  $&\dblue$0$&&\\\hline
 $\alpha_s^5$&\purple&\red&\yellow&\green&\blue${\Gamma}^{(-,5)}_{\rm NLL}  $  &\dblue$0$&\\\hline
 $\alpha_s^6$&\orange&\purple&\red&\yellow&\green&\blue${\Gamma}^{(-,6)}_{\rm NLL}  $  &\dblue$0$\\\hline
     \end{tabular}
\end{center}
     \caption{The signature odd part (imaginary part) of the soft anomalous dimension in the high-energy limit. 
The three-loop coefficient ${\Delta}^{(-,3,0)}$  and ${\Delta}^{(-,3,1)}$ determined from the soft anomalous dimension beyond-dipole correction~\cite{Almelid:2015jia} is given in (\ref{DeltaCoef}).
The NLL anomalous dimension coefficients ${ \Gamma}^{(-,\ell)}_{\rm NLL}$ are given in (\ref{eq:gamma8}); they are known to all orders based on~\cite{Caron-Huot:2017zfo}.  
The unfilled cells at the bottom of the table are yet unknown.}
\label{tabIm}
\end{table}
Turning next to the imaginary part of the amplitude, we observe that non-vanishing corrections beyond the dipole formula have a more prominent role. First, at three loops one has a non-vanishing NNLL imaginary contribution ${\Delta}^{(-,3,1)}$,  as well as a N$^3$LL one ${\Delta}^{(-,3,0)}$, given in eq.~(\ref{DeltaCoef}). Second, Regge cut effects appear here already at the NLL accuracy, i.e. at order $(\alpha_s/\pi)^\ell L^{\ell-1}$, giving rise to a highly non-trivial exponentiation pattern, which was unravelled in~\cite{Caron-Huot:2017zfo}.  These NLL coefficients vanish identically for $\ell=2$ and $3$, but are finite for any $\ell\geq 4$ (see eq.~(\ref{eq:gamma8})). In the following sections we will review the main ideas leading to the computation of these corrections in~\cite{Caron-Huot:2017zfo} using BFKL theory.

\begin{align}
\label{eq:gamma8}
\begin{split}
{ \Gamma}^{(-,1)}_{\rm NLL} &= i \pi  \, \Tsu\,, \qquad
{ \Gamma}^{(-,2)}_{\rm NLL} = 0,\qquad   { \Gamma}^{(-,2)}_{\rm NLL} = 0,\qquad
{ \Gamma}^{(-,4)}_{\rm NLL} = 
- i \pi \, \frac{\zeta_3}{24}\,C_A (C_A - \Tt)^2   \, \Tsu,  \\
{ \Gamma}^{(-,5)}_{\rm NLL} &= 
- i \pi \, \frac{\zeta_4}{128}\,C_A (C_A - \Tt)^3   \, \Tsu, \qquad
{ \Gamma}^{(-,6)}_{\rm NLL} = 
- i \pi \, \frac{\zeta_5}{640}\,C_A (C_A - \Tt)^4   \, \Tsu, \\ 
{ \Gamma}^{(-,7)}_{\rm NLL} &= 
i \pi \frac{1}{720} \bigg[\frac{\zeta_3^2}{16}\,C_A^2 (C_A - \Tt)^4
+\frac{1}{32} \left(\zeta_3^2 - 5 \zeta_6\right) \,C_A (C_A - \Tt)^5  \bigg] \Tsu,  \\ 
{ \Gamma}^{(-,8)}_{\rm NLL} &= 
i \pi \frac{1}{5040} \bigg[\frac{3 \zeta_3\zeta_4}{32}\,C_A^2 (C_A - \Tt)^5
+\frac{3}{64} \left(\zeta_3 \zeta_4 - 3 \zeta_7\right) \,C_A (C_A - \Tt)^6  \bigg] \Tsu.  
\end{split}
\end{align}

\section{BFKL and the two Reggeon cut\label{sec:BFKL_NLL}}

We now turn to discuss the BFKL equation in dimensional regularization, which we set up in order to determine the contribution to the $2\to 2$ amplitude generated by two Reggeized gluon exchange~\cite{Kuraev:1977fs,Balitsky:1978ic,Lipatov:1985uk,Caron-Huot:2013fea,Caron-Huot:2017zfo}. Our formulation is kept general as far as the colour  flow is concerned, but of course it is ultimately relevant only for even $t$-channel representations, which are\footnote{It is well known that for the symmetric octet there are no corrections to the amplitude beyond one loop, and indeed the equation trivialises for ${\mathbf T}_t^2=C_A$.} the singlet and the 27 representation in eq.~(\ref{odd_even_Ms}). 

To strip off the effect of a single Reggized gluon we define a so-called reduced amplitude by $\hat{\cal M}_{ij\to ij} \equiv 
\,e^{-\,{\mathbf T}_t^2\, \alpha_g(t) \, L} \, {\cal M}_{ij\to ij}$, where the $\alpha_g$ is the gluon Regge trajectory, now computed in dimensional regularization including the full $\epsilon$ dependence. We will only need its one-loop expression:
\begin{align}
&\alpha_g(t)=\frac{\alpha_s}{\pi} {\mathbf T}_t^2  \left(\frac{-t}{\mu^2}\right)^{-\epsilon}\frac{B_{0}(\epsilon)}{2\epsilon}
+{\cal O}(\alpha_s^2)\,,
\\
&B_{0}(\epsilon) = e^{\epsilon \gamma_E} \frac{\Gamma^2(1-\epsilon)\Gamma(1+\epsilon)}{\Gamma(1-2\epsilon)} = 1 - \frac{\zeta_2}{2} \epsilon^2 - \frac{7\zeta_3}{3} \epsilon^3 + \ldots\,.
\end{align}

\begin{figure}[b]
\begin{center}
     \includegraphics[width=.56\textwidth]{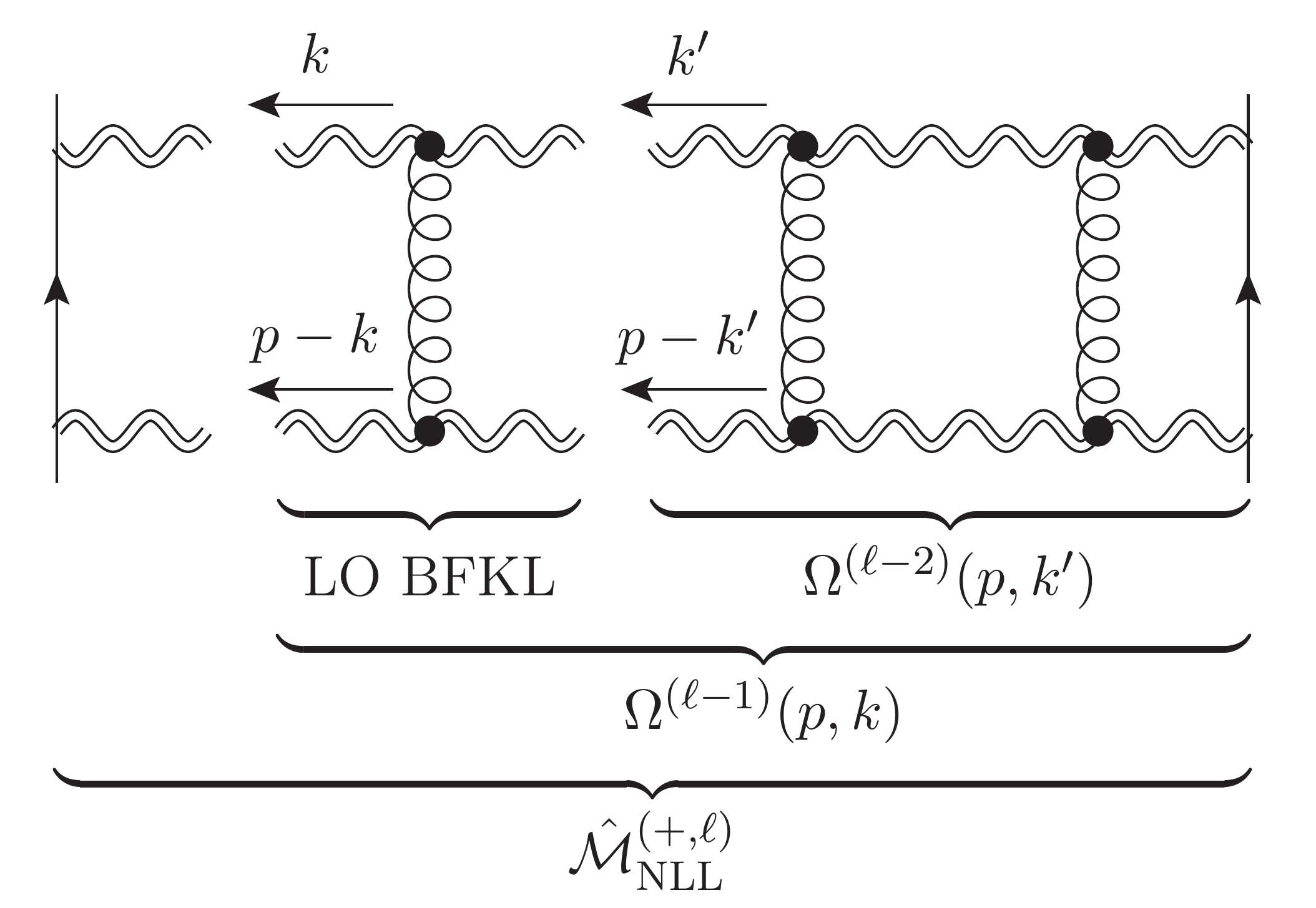}
     \caption{Illustration of the two Reggeized gluon amplitude and the corresponding BFKL wave function. A single application of the BFKL Hamiltonian amounts to adding a rung in the ladder. The amplitude at $\ell$ loops, ${\cal M}^{(\ell)}(p)$, is obtained by integrating over the wavefunction $\Omega^{(\ell-1)}(p,k)$. The latter consists of $\ell-1$ rungs and two off-shell gluons with transverse momenta $k$ and $p-k$.}
     \label{BFKL_Hamiltonian}
\end{center}
     \end{figure}

The reduced amplitude $\hat{\cal M}_{ij\to ij}$ can be expressed as~\cite{Kuraev:1977fs,Balitsky:1978ic,Lipatov:1985uk,Caron-Huot:2013fea,Caron-Huot:2017zfo} 
\begin{align}
\label{Amp}
\hat{{\cal M}}^{(+,\ell)}_{\rm NLL}= -i\pi \frac{B_{0}^\ell}{(\ell-1)!} \int [Dk] \frac{p^2}{k^2(p-k)^2} \Wfl{\ell-1}(p,k) \Tsu \Mtree\,,
\end{align}
where $ \Dk \equiv \frac{\pi}{B_0} 
  \left( \frac{\mu^2}{4\pi e^{-\gamma_E}} \right)^{\epsilon} 
  \frac{\ddeps k}{(2\pi)^{2-2\epsilon}}$ and $ \Wfl{\ell-1}(p,k)$ is the two Reggeized gluon wavefunction at $\ell-1$ loop order, as shown in figure~\ref{BFKL_Hamiltonian}.
The wavefunction admits the BFKL evolution equation
\begin{equation}
\label{evol}
\frac{d}{dL}\Omega(p,k)=\frac{\alpha_s B_0(\epsilon)}{\pi} \hat{H} \Omega(p,k)\,,
\end{equation}
where the Hamiltonian consists of two components $\hat{H} = (2C_A-\Tt) \,\hat H_{\rm i} + (C_A-\Tt) \, \hat H_{\rm m}$ with
\begin{subequations}
\label{Him}
\begin{align}
\hat H_{\rm i}\,\Psi(p,k) &=\int [{\mathrm D}k'] \, f(p,k,k') \Big[\Psi(p,k')-\Psi(p,k)\Big]\,,\\
\hat H_{\rm m}\,\Psi(p,k) &= J(p,k)\, \Psi(p,k)\,,
\end{align}
\end{subequations}
where
\begin{equation}
\label{fk}
f(p,k,k') \equiv \frac{k^2}{k'^2(k - k')^2}
+\frac{(p-k)^2}{(p-k')^2(k - k')^2}
-\frac{p^2}{k'^2(p- k')^2}\,,
\end{equation}
and
\begin{align}
\label{J}
J(p,k) &= \frac{1}{2\epsilon} + \int [{\mathrm D}k'] \, f(p,k,k') = \frac{1}{2\epsilon} \left[2- \left(\frac{p^2}{k^2}\right)^{\epsilon}-\left(\frac{p^2}{(p-k)^2}\right)^{\epsilon}\right].
\end{align}
While the exact solution is unknown, the equation can be solved iteratively 
\begin{equation}
\Wf(p,k) = \sum_{\ell=1}^{\infty} \left( \frac{\alpha_s}{\pi}B_{0} \right)^{\ell}  \frac{L^{\ell-1}}{(\ell -1)!} \, \Wfl{\ell-1}(p,k),\qquad \Omega^{(\ell-1)}(p,k) =
\hat{H} \,\Omega^{(\ell-2)}(p,k),
\end{equation}
and we get
\begin{eqnarray}
\Omega^{(0)}(p,k) &=& 1\\\nn
\Omega^{(1)}(p,k) &=& (C_A-{\mathbf T}_t^2)  J(p,k)\\\nn
\Omega^{(2)}(p,k) &=& (C_A-{\mathbf T}_t^2)^2 J^2(p,k)  + (C_A-2{\mathbf T}_t^2) (C_A-{\mathbf T}_t^2) \int [{\mathrm D}k'] f(p,k,k') \left[ J(p,k') - J(p,k) \right]\,,
\end{eqnarray}
and so on. Direct integration of these iterative formulae is possible for the first few orders (we computed the amplitude (\ref{Amp}) this way, as an expansion in $\epsilon$, through five loops), but becomes intricate at higher orders, where we resort to more sophisticated techniques elaborated in sections \ref{sec:BFKL_soft} and \ref{sec:BFKL_2d}. 

\section{The soft approximation\label{sec:BFKL_soft}}

The first crucial observation of \cite{Caron-Huot:2017zfo} is that the wavefunction $\Omega^{(\ell)}(p,k)$ is finite, to any order.  This can be deduced directly from the form of the Hamiltonian (\ref{Him}).
It implies that all infrared singularities in the NLL amplitude are generated in the final integration over $k$ in eq.~(\ref{Amp}), from the soft regions where $k\ll p$ or where $p-k\ll p$. We may consider the former and recover the latter using symmetry.

The next observation is that BFKL evolution closes in the soft approximation, $k\ll p$. To see this it suffices to consider a single evolution step of the wavefunction generated by the Hamiltonian (\ref{Him}). To understand the action of $\hat H_{\rm i}$ one needs to identify the $k'$ momentum regions that dominate the integral when $k\ll p$ .  A priori there are two potential regions that could contribute, $k\sim k'\ll p$ and $k\ll k'\sim p$, however since in the latter $f(p,k,k')$ in (\ref{fk}) vanishes, only the former contributes, and thus $\Omega^{(\ell)}(p,k)$ for $k\ll p$ is fully determined by $\Omega^{(\ell-1)}(p,k')$ with $k'\ll p$. Applying this argument iteratively we conclude that the soft limit of the wavefunction is fully determined by the configuration where the entire side rail of the ladder is soft. Upon expanding the evolution equation for $k\sim k'\ll p$ we get
\begin{eqnarray}
\label{softevol}
\Omega^{(\ell-1)}_s(p,k)&=&\hat H_s \, \Omega^{(\ell-2)}_s(p,k)\,,
\\
\hat H_s \Psi(p,k)&=& (2C_A-{\mathbf T}_t)\int[{\mathrm D}k']
\frac{2(k\cdot k')}{k'^2(k - k')^2} \, \Big[ \Psi(p,k')-\Psi(p,k)\Big] +\, (C_A-{\mathbf T})\, J_s(p,k) \, \Psi(p,k)\,,\nn
\end{eqnarray}
with $J_s(p,k) = \frac{1}{2\epsilon} \left[1-\left({p^2}/{k^2}\right)^{\epsilon}\,\right]$. 
The next observation is that solving the evolution in the soft limit, eq.~(\ref{softevol}), is drastically simpler than solving the general equation described in the previous section, because it reduces a two-scale problem, $k^2$ and $(p-k)^2$, into a single-scale problem. The integrals over the kernel are then simple to perform:
\begin{align}
\begin{split}
&\int [{\mathrm D}k'] \, \frac{2(k\cdot k')}{k'^2(k - k')^2} \, 
\left(\frac{p^2}{k'^2}\right)^{n \epsilon} =
-\frac{1}{2\epsilon}\frac{B_{n}(\epsilon)}{B_{0}(\epsilon)}\left(\frac{p^2}{k^2}\right)^{(n+1)\epsilon}
\\
&\text{with} \qquad B_{n}(\epsilon)= e^{\epsilon\gamma_{\rm E}}  \frac{\Gamma(1-\epsilon)}{\Gamma(1+n\epsilon)}
\frac{\Gamma(1+\epsilon + n\epsilon) \Gamma(1-\epsilon - n\epsilon)}{\Gamma(1-2 \epsilon - n\epsilon)}\,,
\end{split}
\end{align}
and the solution for the wavefunction at any order can be simply written as a polynomial in $\xi \equiv (p^2/k^2)^\epsilon$:
\begin{align}
\label{OmegaSoft}
\Omega^{(\ell-1)}(p,k) =
\frac{(C_A-{\mathbf T}_t)^{\ell-1}}{(2\epsilon)^{\ell-1}} \sum_{n=0}^{\ell-1} 
(-1)^n 
\left(\begin{array}{c}{\ell-1}\\{n}\end{array} \right)
\left(\frac{p^2}{k^2}\right)^{n\epsilon} 
\prod_{m=0}^{n-1} \left\{1 - \hat{B}_{m}(\epsilon) \frac{2C_A-{\mathbf T}_t}{C_A-{\mathbf T}_t}\right\}\,,
\end{align}
where $\hat B_n(\epsilon) = 1- \frac{B_n(\epsilon)}{B_0(\epsilon)}
\,=\,
2 n (2 + n) \zeta_3 \epsilon^3 
+ 3 n (2 + n) \zeta_4 \epsilon^4 
+\ldots.
$
The first few orders read:
\begin{align}
\begin{split}
\Omega_{0}(\xi) =& 1, \qquad\quad 
\Omega_{1}(\xi) = \frac{(C_A-{\mathbf T}_t)}{2\epsilon} 
\Big(1-\xi\Big), \\ 
\Omega_{2}(\xi) =& \frac{(C_A-{\mathbf T}_t)^2}{(2\epsilon)^2} 
\bigg\{1-2\xi+ \xi^2\left[ 1- \hat{B}_{1}(\epsilon)\frac{2C_A-{\mathbf T}_t}{C_A-{\mathbf T}_t}\right]\bigg\}\,,\\ 
\Omega_{3}(\xi) =& \frac{(C_A-{\mathbf T}_t)^3}{(2\epsilon)^3} 
\bigg\{1-3\xi + 3\xi^2\left[ 1- \hat{B}_{1}(\epsilon)\frac{2C_A-{\mathbf T}_t}{C_A-{\mathbf T}_t}\right] \\ 
&\hspace{25mm}-\,\xi^3\left[ 1- \hat{B}_{1}(\epsilon)\frac{2C_A-{\mathbf T}_t}{C_A-{\mathbf T}_t}\right]
\left[ 1- \hat{B}_{2}(\epsilon)\frac{2C_A-{\mathbf T}_t}{C_A-{\mathbf T}_t}\right]\bigg\}\,.
\end{split}
\end{align}
Performing the final integration over $k$ according to (\ref{Amp}) we obtain all the infrared-singular contributions to the NLL amplitude at any loop order $\ell$. Remarkably, the result  can be resummed into a closed-form expression:
\begin{align}
\label{softAmplRes}
\left.\hat{\cal M}_{\rm NLL}^{(+)}\right|_s = \frac{i\pi}{L(C_A-{\mathbf T}^2_t)} 
 \left( 1 - R(\epsilon) \frac{C_A}{C_A -{\mathbf T}_t^2} \right)^{-1} 
  \left[ \exp\left\{\frac{B_0(\epsilon)}{2\epsilon} 
  \frac{\alpha_s}{\pi} L (C_A-{\mathbf T}_t)\right\} - 1 \right] {\mathbf T}^2_{s-u} \, {\cal M}^{({\rm tree})}\, 
\end{align}
up to  ${\cal O}(\epsilon^0)$ terms,  where we defined 
\begin{eqnarray}
\label{Reps}
R(\epsilon) \equiv \frac{B_0(\epsilon)}{B_{-1}(\epsilon)} -1 &=&
\frac{\Gamma^{3}(1-\epsilon)\Gamma(1+\epsilon)}{\Gamma(1-2\epsilon)} -1  \\
&=& -2\zeta_3 \, \epsilon^3 -3\zeta_4 \, \epsilon^4 -6\zeta_5 \epsilon^5
-\left(10 \zeta_6-2\zeta^2_3 \right) \epsilon^6 + {\cal O}(\epsilon^7).\nn
\end{eqnarray}
This result is consistent with the exponentiation of infrared singularities, yielding the NLL contributions to the soft anomalous dimensions already summarised in eq.~(\ref{eq:gamma8}) above.

\section{Hard contributions to BFKL evolution using two-dimensional transverse space\label{sec:BFKL_2d}}

Having determined the singularities, the next challenge is to determine ${\cal O}(\epsilon^0)$ finite terms in the amplitude at NLL. The key to doing this~\cite{Caron-Huot:2019_TBP} is again the fact that the wavefunction itself is finite, and hence can be computed consistently in two transverse dimensions, i.e. setting $\epsilon=0$. Of course, doing this we should ultimately address the challenge of determining the amplitude from the wavefunction, noting that the integral in (\ref{Amp}) requires (dimensional) regularization.  This problem was solved in~\cite{Caron-Huot:2019_TBP} by elegantly combining the soft limit discussed above with the two-dimensional calculation, as we briefly explain at the end of this section.  

Let us focus first on the calculation of the wavefunction in exactly two (transverse) dimensions. Representing the two-dimensional momentum 
vectors $k$, $k'$ and~$p$ as complex numbers, 
$k = k_x + i k_y,$ \, $k' = k_x' + i k_y'$ and $p = p_x + i p_y$,
we may change variables in the BFKL equation as follows:
\begin{equation}
  \label{eq:zwdef} \frac{k_x + i k_y}{p_x + i p_y} = 
  \frac{z}{z-1} \qquad \text{and} \qquad \frac{k_x' + i k_y'}{p_x + i p_y} 
  = \frac{w}{w-1}.
\end{equation}
Since the wavefunction is a function of Lorentz scalars 
(i.e.\ squares of momenta) it is symmetric under the 
exchange $z \leftrightarrow \zb$ with $\zb$ the complex 
conjugate of $z$.  In the new variables the kernel \eqref{fk} 
reads
\begin{equation}
  \label{eq:f2d} p^2 f(p,k,k') \longrightarrow (1-w)^2 (1-\wb)^2 K(w,\wb,z,\zb),
\end{equation}
where
\begin{equation}
  \label{eq:K2d} K(w,\wb,z,\zb) = \frac{z\wb+w\zb}{w\wb(z-w)(\zb-\wb)} 
  = \frac{1}{\wb(z-w)} + \frac{2}{(z-w)(\zb-\wb)} + \frac{1}{w(\zb-\wb)}. 
\end{equation}
Furthermore, in the limit $\epsilon \to 0$, $J(p,k)$ of eq.~\eqref{J} and the measure becomes, respectively, 
\begin{equation}
  \label{eq:j2d} J(p,k) \longrightarrow j(z,\zb) \equiv 
  \frac12 \log \left[ \frac{z}{(1-z)^2} \frac{\zb}{(1-\zb)^2} \right],
\qquad\quad
 \frac{\dd^2 k'}{p^2} \longrightarrow 
  \frac{\dd^2 w}{(1-w)^2 (1-\wb)^2}.
\end{equation}
We may thus formulate the iterative solution of the BFKL equation in exactly two dimensions as $\Wtdl{\ell}(z,\zb) = \Htd \Wtdl{\ell-1}(z,\zb)$,
where the Hamiltonian is
\begin{equation}
 \Htd \psi(z,\zb) = 
  C_1 \Hitd \psi(z,\zb) 
 +  C_2 \Hmtd \psi(z,\zb).
\end{equation}
where the colour factors are denoted by $C_1=\Cone$ and $C_2= \Ctwo$ and
where
\begin{align}
  \label{eq:hi2} 
\begin{split}\Hitd \psi(z,\zb) &= \frac{1}{4\pi} \int \dd^2 w K(w,\wb,z,\zb) \left[ \psi(w,\wb) - \psi(z,\zb) \right], \\
\Hmtd \psi(z,\zb) &= j(z,\zb) \psi(z,\zb),
\end{split}
\end{align}
with $\Wtdl{0}(z,\zb) = \Wfl{0}(p,k) = 1$. We note that the two dimensional Hamiltonian admits two symmetries $z\leftrightarrow \zb$ and $z\leftrightarrow 1/z$, the latter corresponding to the interchange of the two Reggeons.

The next crucial observation in~\cite{Caron-Huot:2019_TBP}, which greatly simplifies the iterative solution, is that the wavefunction, at any given order, can be expressed in terms of single-valued harmonic polylogarithms (SVHPLs), the same class of functions we have already encountered in the discussion of the three-loop soft anomalous dimension (for further details about these functions, see~\cite{Brown:2004ugm,Brown:2013gia,Schnetz:2013hqa,Pennington:2012zj,Dixon:2012yy,DelDuca:2013lma}).  The single-valuedness is expected in the present context since branch cuts are physically inadmissible in the Euclidean two-dimensional transverse space.
Furthermore, the structure Hamiltonian guarantees that the wavefunction at order $\ell$ is a pure function of uniform weight $\ell$.
Practically, the restriction to SVHPLs has far reaching consequences: the result of applying the Hamiltonian $\hat{H}_{{\rm 2d},{\rm i}}$ in (\ref{eq:hi2}) to any linear combination of SVHPLs ${\cal L}_w(z,\zb)$ (where the word $w$ corresponds to a set of 0 and 1 indices) can be deduced from the following set of differential equations:
\begin{align}
\label{diffeqs}
\begin{split}
\frac{d}{dz} \hat{H}_{{\rm 2d},{\rm i}} \mathcal{L}_{0,\sigma}(z,\bar{z}) &= \frac{\hat{H}_{{\rm 2d},{\rm i}} \mathcal{L}_{\sigma}(z,\bar{z})}{z}  \,,
\\
\frac{d}{dz} \hat{H}_{{\rm 2d},{\rm i}} \mathcal{L}_{1,\sigma}(z,\bar{z})  &= \frac{\hat{H}_{{\rm 2d},{\rm i}} \mathcal{L}_{\sigma}(z,\bar{z})}{1-z}   - \frac14 \frac{\mathcal{L}_{1,\sigma}(z,\bar{z})}{z}  
\\
  &\hspace{-5mm} - \frac14 \frac{\mathcal{L}_{0,\sigma}(z,\bar{z})  + 2\mathcal{L}_{1,\sigma}(z,\bar{z}) - [\mathcal{L}_{0,\sigma}(w,{\bar{w}}) + \mathcal{L}_{1,\sigma}(w,{\bar{w}})]_{w,{\bar{w}} \rightarrow \infty}}{1-z}\,,
\end{split}
\end{align}
which represent the fact that certain logarithmic derivatives (such as $z\frac{d}{dz}$) commute with the Hamiltonian up to contact terms.  The latter arise from the fact that $w$ and $\bar{w}$ are not independent when their derivatives act on singular terms, e.g. $\frac{d}{dw} \frac{1}{\wb-c} = \pi \delta^2(w-c)$. These contact terms give rise to the terms involving the $w,\wb\to\infty$ limit in the second equation in~(\ref{diffeqs}).
The differential equations (\ref{diffeqs}) can be integrated to iteratively build the wavefunction at any order, using the soft limit $z,\zb\to 0$ as boundary data; the latter can be obtained from (\ref{OmegaSoft}) upon taking $\epsilon=0$. This procedure has been automated in~\cite{Caron-Huot:2019_TBP}, and the resulting wavefunction at the first few orders reads:
\begin{eqnarray}
\label{Omega2d}
\Wtdl{1} &=& \frac{1}{2} C_2 \left(\El_0+2 \El_1\right) \\ \nn
	\Wtdl{2} &=& \frac{1}{2} C_2^2 \left(\El_{0,0}+2 \El_{0,1}
	+2 \El_{1,0}+4 \El_{1,1}\right)+\frac{1}{4} C_1 C_2 \left(-\El_{0,1}
	-\El_{1,0}-2 \El_{1,1}\right) \\ \nn
	\Wtdl{3} &=&
\frac{3}{4} C_2^3 \left(\El_{0,0,0}+2 \El_{0,0,1}
	+2 \El_{0,1,0}+4 \El_{0,1,1}+2 \El_{1,0,0}
 +4 \El_{1,0,1}+4 \El_{1,1,0}
	+8 \El_{1,1,1}\right)
 \\\nn	&&\hspace{4mm}+
 \frac{1}{4} C_1 C_2^2 \left(2 \zeta_3 -2 \El_{0,0,1}
	-3 \El_{0,1,0}-7 \El_{0,1,1}-2 \El_{1,0,0}
	-7 \El_{1,0,1}-7 \El_{1,1,0}
-14 \El_{1,1,1}\right)
 \\\nn &&\hspace{4mm}
	+\frac{1}{16} C_1^2 C_2 \left(\El_{0,0,1}
	+2 \El_{0,1,0}+4 \El_{0,1,1} 
+\El_{1,0,0}+4 \El_{1,0,1}
	+4 \El_{1,1,0}+8 \El_{1,1,1}\right)\,.
\end{eqnarray}

Having determined the two-dimensional wavefunction to any required order, let us return to the question raised early on in this section, namely how to obtain the $2\to 2$ amplitude itself despite the singular nature of (\ref{Amp}), which requires a dimensionally-regularized wavefunction. In~\cite{Caron-Huot:2019_TBP} this was done by separating the full wavefunction (defined in dimensional regularization) into hard and soft components $ \Omega(p,k) = \Omega_\text{hard}(p,k) + \Omega_\text{soft}(p,k)$. For $\Omega_\text{soft}(p,k)$ we use the symmetrized version of the dimensionally-regularized  solution in eq.~(\ref{OmegaSoft}), where $(p^2/k^2)^\epsilon$ is replaced by 
$\left(\frac{(p^2)^2}{k^2(p-k)^2}\right)^{\epsilon/2}$, which captures both soft limits and furthermore, admits an $\epsilon$ expansion in terms of SVHPLs, and hence its two-dimensional limit reproduces exactly all nonvanishing terms in the $z\to 0$ and $z\to \infty$ limits of the two-dimensional solution $\Omega^{(\rm 2d)}(z,\bar{z})$ of (\ref{Omega2d}). Thus we are able to isolate the two-dimensional limit of the hard wavefunction:
\begin{align}
 \Omega_\text{hard}^{({\rm 2d})}(z,\bar{z})\,\equiv \,
\lim_{\epsilon \to 0} \Omega_{\text{hard}} = \Omega^{(\rm 2d)}(z,\bar{z}) - \Omega_{\text{soft}}^{({\rm 2d})}(z,\bar{z})
\end{align}
Given that all the singularities arise from the soft component of the wavefunction, we then obtain the NLL amplitude (\ref{Amp}) through finite terms by 
\begin{align}
\label{AmpComb}
 \hat{\cal M}^{(+,\,\text{NLL})}_{ij\to ij}\left(\frac{s}{-t}\right) = -i\pi \left[ \int [{\rm D}k] \frac{p^2}{k^2(p-k)^2} \Omega_{\rm soft}(p,k) + \frac{1}{4\pi} \int \frac{d^2 z}{z\bar{z}} \Omega_{\rm hard}^{({\rm 2d})}(z,\bar{z}) \right] {\mathbf T}^2_{s-u} {\cal M}^{({\rm tree})}_{ij\to ij}\,,
\end{align}
where we integrate the $\epsilon$-dependent soft wavefunction in dimensional regularization and the hard wavefunction over the two-dimensional measure: given that the latter vanishes, by construction, in the soft limits, the two dimensional integral converges. The exact ${\cal O}(\epsilon^0)$ terms in the amplitude are recovered in the sum of the two integrals in the square brackets.

\section{The imaginary part of the $2\to 2$ amplitude: results and analysis\label{sec:results}}  

\begin{figure}[t]
  \centering
  \includegraphics{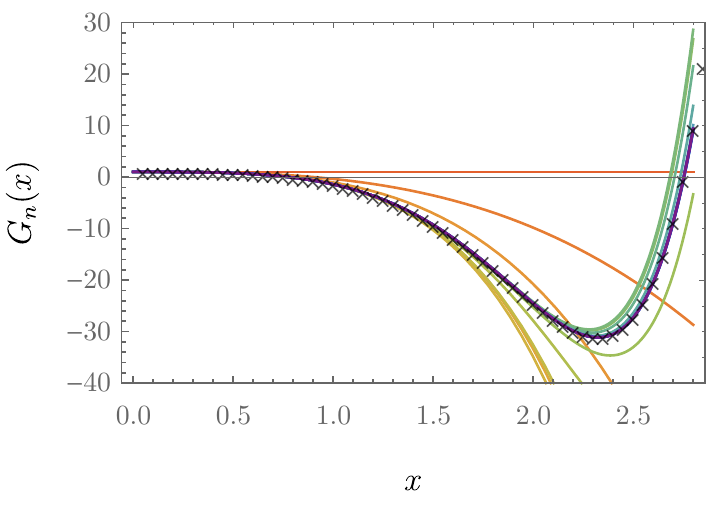}\hspace{9pt} 
  \includegraphics{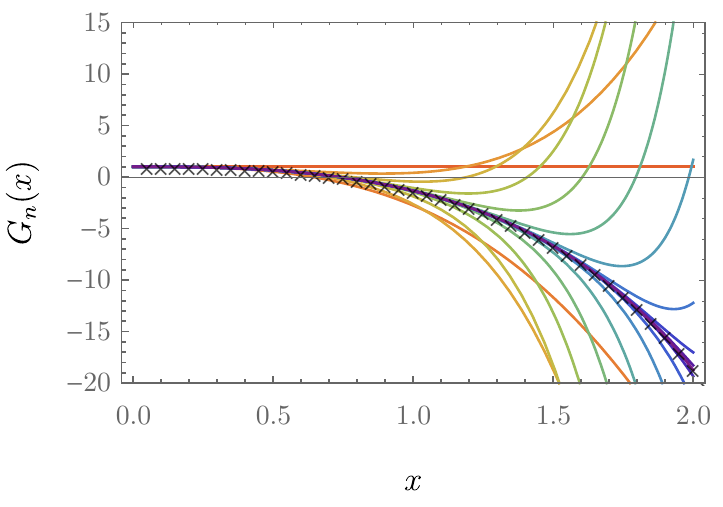}
  \caption{Partial sums $G_n(x) = \sum_{\ell=1}^{n} 
  G^{(\ell)} x^{\ell-1}$ for $n=1,\dots,22$ (rainbow, red 
  through violet) and numerical results for $G(x)$ (black 
  crosses). The horizontal axis $x$ represents $\frac{\alpha_s}{\pi}L$. The plot illustrates convergence in that increasing 
  the order~$n$ extends the range of $x$ for which the partial 
  sum matches the numerical result. 
The figure shows 
  the singlet (left) as well as the $27$ 
  exchange (right).}
  \label{fig:Glsums}
\end{figure}

Upon performing the integrals in (\ref{AmpComb}) we obtain the full NLL $2\to2$ amplitude, which is the leading tower of logarithms in the imaginary part of the amplitude. 
\begin{align}
\label{AmplRes}
\begin{split}
 \hat{\cal M}^{(1)}_{ij\to ij} =&i\pi \frac{1}{2 \epsilon }{\mathbf T}^2_{s-u} {\cal M}^{({\rm tree})}_{ij\to ij}
\\
\hat{\cal M}^{(2)}_{ij\to ij} =&i\pi
C_2 \bigg[
\frac{1}{8 \epsilon ^2}
-\frac{ \zeta (2)}{8} 
\bigg]
{\mathbf T}^2_{s-u} {\cal M}^{({\rm tree})}_{ij\to ij}
\\
\hat{\cal M}^{(3)}_{ij\to ij} =&i\pi C_2^2 \bigg[\frac{1}{48 \epsilon ^3} -\frac{ \zeta (2)}{32 \epsilon } -\frac{29}{48}   \zeta (3)\bigg]{\mathbf T}^2_{s-u} {\cal M}^{({\rm tree})}_{ij\to ij}
\\
\hat{\cal M}^{(4)}_{ij\to ij} =&i\pi C_2^2 \bigg[ \frac{C_2}{384 \epsilon ^4}-\frac{C_2 \zeta (2)}{192 \epsilon ^2} -\left(\frac{7C_2 }{288} +\frac{C_A }{192}  \right)
\frac{\zeta (3)}{\epsilon }-\frac{C_2 \zeta (4)}{48}-\frac{C_A \zeta (4)}{128}\bigg]{\mathbf T}^2_{s-u} {\cal M}^{({\rm tree})}_{ij\to ij}
\\
\hat{\cal M}^{(5)}_{ij\to ij} =&i\pi C_2^2 \bigg[\frac{C_2^2}{3840 \epsilon ^5}-\frac{C_2^2 \zeta (2)}{1536 \epsilon ^3}+
\left({-\frac{7 C_2^2 }{2304}-\frac{C_2 C_A }{1920}}\right)\frac{\zeta (3)}{\epsilon ^2}+\left({-\frac{9 C_2^2 }{4096}-\frac{C_2 C_A}{1280}}\right)\frac{\zeta (4)}{\epsilon }
\\&\hspace*{-20pt}+C_2^2 \left(\frac{35 \zeta (2) \zeta (3)}{4608}-\frac{293 \zeta (5)}{1280}\right)+C_2 C_A \left(\frac{\zeta (2) \zeta (3)}{768} +\frac{253 \zeta (5)}{1920}\right)-\frac{\zeta (5) }{48} C_A^2 \bigg]{\mathbf T}^2_{s-u} {\cal M}^{({\rm tree})}_{ij\to ij}
\end{split}
\end{align}
In~\cite{Caron-Huot:2019_TBP} we computed these expressions through to order $\ell=13$ loops, and higher orders can be computed by the same algorithm, although the expressions and the evaluation time get long. The result admits uniform transcendental weight. At high loop orders, ${\cal O}(\epsilon^0)$ terms feature multiple zeta values (the first is $\zeta_{5,3,3}$ at 11 loops), but only of the type that originates in SVHPLs (single zeta values appear also through the soft limit, which is not restricted to the single-valued class). As discussed above,  the singularities can be resummed into (\ref{softAmplRes}). They also exponentiate in terms of the soft anomalous dimension whose first few orders were quoted in (\ref{eq:gamma8}). The NLL anomalous dimension can be written as
${\Gamma}^{(-,\ell)}_{\rm NLL} = i\pi \,G^{(\ell)} \,\Tsu$
where the coefficients are generated by
\begin{equation}
\label{Gl}
G^{(\ell)} \equiv \frac{1}{(\ell-1)!}\left[ \frac{(C_A-\Tt)}{2}\right]^{\ell-1} 
\left.\left( 1 - R(\epsilon) \frac{C_A}{C_A -\Tt} \right)^{-1}\right\vert_{\epsilon^{\ell-1}} \,\,,
\end{equation}
where $R(\epsilon)$ is given by (\ref{Reps}) and $\vert_{\epsilon^{\ell-1}}$ indicates that one should extract the coefficient of $\epsilon^{\ell-1}$. 
The NLL soft anomalous dimension itself can be resummed~\cite{Caron-Huot:2017zfo}
\begin{equation} \label{GammaNLL3}
{\Gamma}_{\rm NLL}^{(-)} = i\pi \frac{\alpha_s}{\pi} 
\,G\left(\frac{\alpha_s}{\pi}L\right)\Tsu\,,
\end{equation}
and interestingly, $G(x) = \sum_{\ell=1}^\infty x^{\ell-1} G^{(\ell)}$ is an entire function, admitting an infinite radius of convergence. Remarkably, we are therefore able to compute the NLL soft anomalous dimension at any value of the effective coupling~\cite{Caron-Huot:2017zfo}, including $\frac{\alpha_s}{\pi}L\gg1$. The good convergence properties are illustrated in figure.~\ref{fig:Glsums}.

\begin{figure}[htb]
  \centering
  \includegraphics[width=0.45\textwidth]{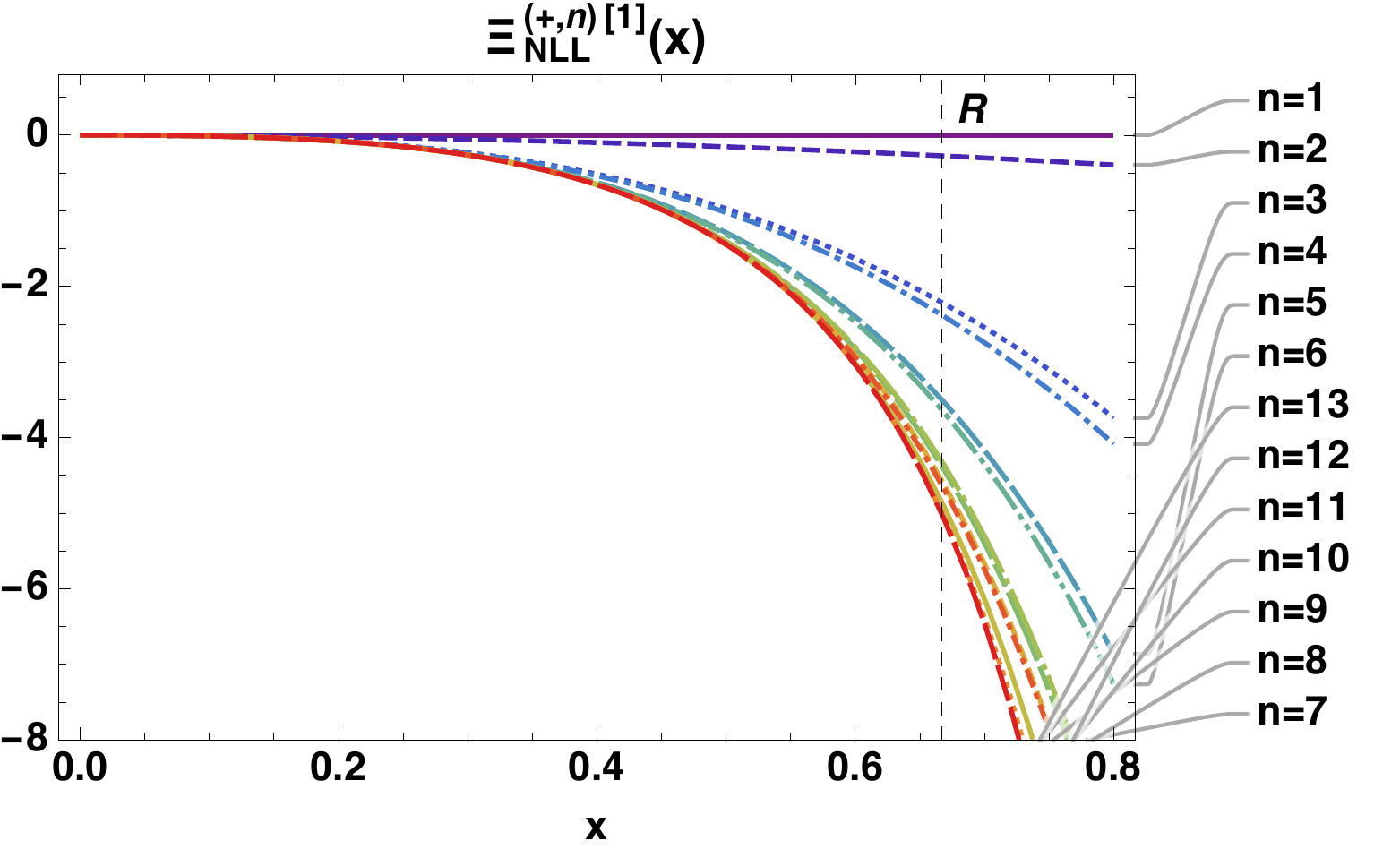}  
  \includegraphics[width=0.45\textwidth]{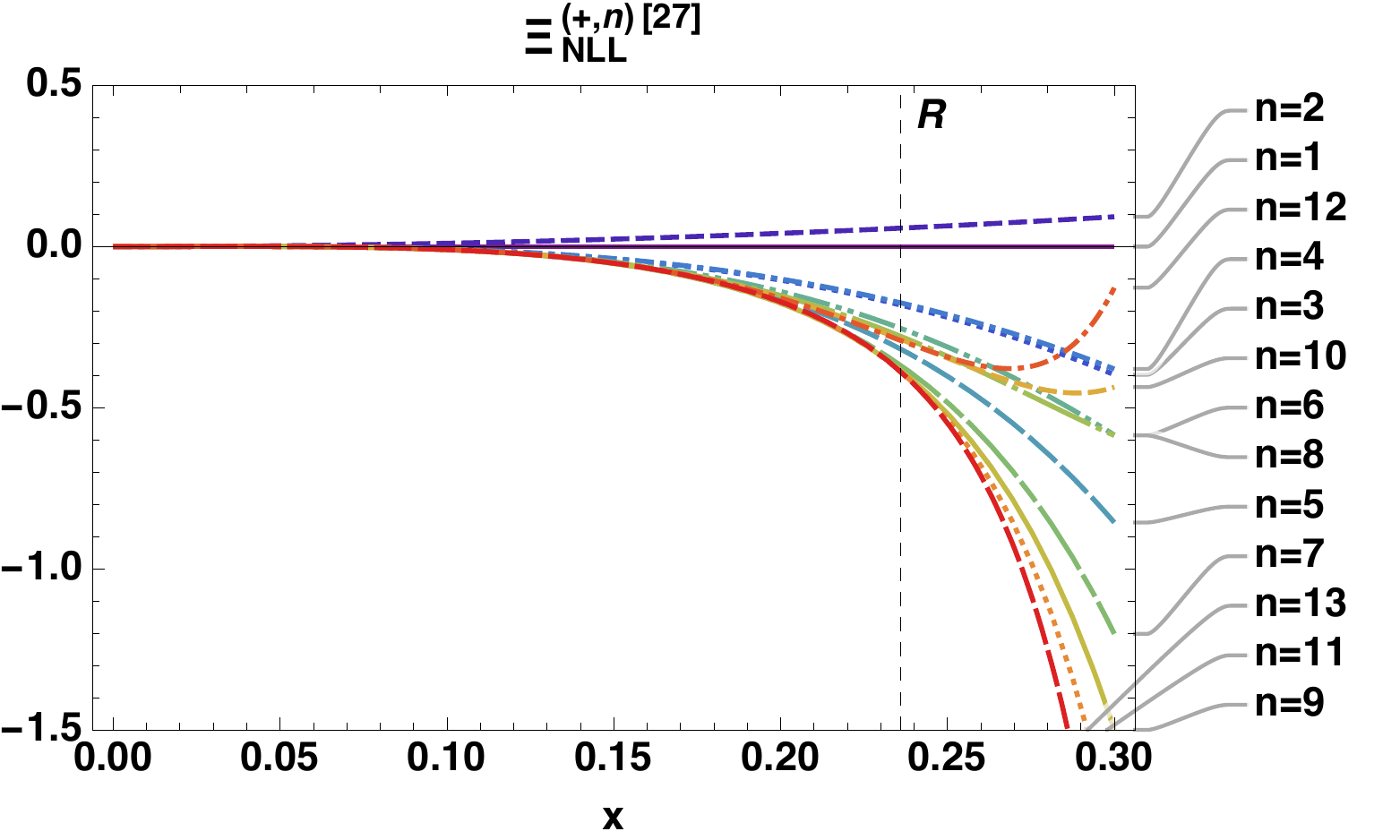}
  \caption{Partial sums of the amplitude coefficients $\Xi_{\rm NLL}^{(+,\ell)}$, 
  up to 13th order, for the singlet (upper plot) and the 27 colour representation (lower plot). The horizontal axis $x$ represents $\frac{\alpha_s}{\pi}L$. The dashed  
  vertical line represents the radius of convergence, $R$, determined by the  pole closest to $x = 0$, using Pad\'{e} approximants.}
  \label{Radius-Full}
\end{figure}

The finite, ${\cal O}(\epsilon^0)$ contribution to the NLL amplitude (\ref{AmplRes}) can be written as
\begin{equation}
\hat{{\cal M}}^{(+)}_{\rm NLL}  = \frac{i \pi}{L} \, \Xi_{\rm NLL}^{(+)} \, \Tsu {\cal M}_{\rm tree}\,.
\end{equation}
It is not yet known how to resum $\Xi_{\rm NLL}^{(+)}$. It is clear however that this resummation would not involve only $\Gamma$ functions, because they contain (single valued) multiple zeta values. Numerically, for the relevant representations for $N_c=3$, the singlet (${\mathbf T}_{t}^2 \, {\cal M}^{[1]} = 0$) and the 27 representation (${\mathbf T}_{t}^2 \, {\cal M}^{[27]}  = 2(N_c+1)\, {\cal M}^{[27]}=8\, {\cal M}^{[27]}$),  $\Xi_{\rm NLL}^{(+)}$ evaluates to as follows:
\begin{eqnarray} 
\label{Xi1} 
\Xi_{\rm NLL}^{(+)[1]} &=& -0.6169 \, x^2 - 6.536 \, x^3 
- 0.8371 \, x^4 - 8.483 \, x^5 - 1.529 \, x^6 - 12.67 \, x^7 + 1.610 \, x^8 \\
&&\hspace*{5pt}-\, 20.62 \,  x^9 + 16.48\,  x^{10} - 35.98 \, x^{11} + 46.07 \, x^{12} 
 - 74.04 \, x^{13} + {\mathcal{O}}(x^{14}),\nn \\
\label{Xi27} 
\Xi_{\rm NLL}^{(+)[27]} &=& 1.028 \, x^2 - 18.16 \, x^3 
+ 2.184 \, x^4 - 196.0 \, x^5 + 372.3 \, x^6 - 2821 \, x^7  + 9382 \,  x^8 \\ 
&&\hspace*{5pt} -\, 46494 \, x^9  + 180397 \, x^{10}  - 797524 \, x^{11} 
+ 3.239 \times 10^{6} \,  x^{12} - 1.374 \times 10^{7} \, x^{13} + {\mathcal{O}}(x^{14}). \nn
\end{eqnarray}
Here we have sufficiently high orders to study the convergence properties of the series. In~\cite{Caron-Huot:2019_TBP} this was done in detail using Pad\'{e} approximants, concluding that the series has a finite radius of convergence.
This is illustrated in figure~\ref{Radius-Full}.
Specifically, we find asymptotic geometric progression with the powers of 
$-\frac12 C_2\frac{\alpha_s}{\pi}L=-\frac32 x$ and $({C_2}- \frac{3}{8}{C_1})  \frac{\alpha_s}{\pi}L=-\frac{17}{4}x$
for the singlet and the 27 representations, respectively. 
 In both cases the series displays sign-oscillations, indicating that once resummed, it could be extrapolated beyond the convergence radius.

\section{Conclusions\label{sec:conclusions}}

In the first part of the talk we have illustrated the complementarity of the high-energy limit and infrared factorization as avenues in studying gauge theory amplitudes. We have done that by considering the two limits sequentially in different orders, ultimately equating the high-energy limit of the soft anomalous dimension to the soft limit of BFKL. Combining the various approaches, the state-of-the-art knowledge of the soft anomalous dimension in the high-energy limit is presented in tables~\ref{tabReal} and~\ref{tabIm}.

In the second part of the talk we demonstrated that rapidity evolution equations can be efficiently used to compute partonic scattering amplitudes to high loop orders. Specifically, summarising the main results of refs.~\cite{Caron-Huot:2017zfo,Caron-Huot:2019_TBP}, we focussed on the leading tower of logarithms in the imaginary part of $2\to 2$ scattering amplitudes, which is generated by the exchange of a pair of Reggeized gluons.  This NLL amplitude is determined by the leading-order BFKL equation, and in QCD it  receives contributions from the singlet and the 27 representations. 

Using the fact that the BFKL wavefunction is finite, we used a combination of techniques to solve the equation iteratively for a general colour flow.
We first used the soft approximation in order to determine the singularities of the amplitude. We showed that these can be resummed in a closed form involving only gamma functions, and also exponentiate in terms of the soft anomalous dimension. At this logarithmic accuracy the latter is found to have an infinite radius of convergence.
We then used an iterative solution of the BFKL equation in two transverse dimensions in order to determine all finite corrections to the amplitude. The calculation is greatly simplified by the fact that the two-dimensional two-Reggon wavefucntion is expressible in terms of single-valued HPLs.
Finite corrections to the amplitude display a more complicated pattern, and cannot be resummed in terms of gamma functions. Indeed, in contrast to the singularities, they involve multiple zeta values (of the single-valued type).                      
We also find that the finite part of the NLL amplitude has a finite radius of convergence in $\frac{\alpha_s(-t)}{\pi}\ln\frac{s}{-t}$, with asymptotically sign-oscillating coefficients. This suggests that its resummed expression, once obtained, could be extrapolated to high-energies (or large coupling) beyond the convergence radius.

\bibliographystyle{JHEP}
\bibliography{main}

\end{document}